\documentclass[11pt,reqno]{amsart}

\newtheorem{prop}{Proposition}[section]
\newtheorem{thm}{Theorem}[section]

\newtheorem{rmk}{Remark}[section]

\newcommand{\mysection}[1]{\section{#1}\setcounter{equation}{0}}

\newfont{\bb}{msbm10 at 12pt}

\def\pf{{\textit {Proof :} }}

\def\R{\hbox{\bb R}}

\def\N{\mathcal N}

\newcommand{\bal}{\begin{align}}      \newcommand{\eal}{\end{align}}
\newcommand{\ba}{\begin{array}}      \newcommand{\ea}{\end{array}}
\newcommand{\bc}{\begin{center}}     \newcommand{\ec}{\end{center}}
\newcommand{\be}{\begin{enumerate}}  \newcommand{\ee}{\end{enumerate}}
\newcommand{\beq}{\begin{eqnarray}}  \newcommand{\eeq}{\end{eqnarray}}
\newcommand{\beQ}{\begin{eqnarray*}} \newcommand{\eeQ}{\end{eqnarray*}}
\newcommand{\bi}{\begin{itemize}}    \newcommand{\ei}{\end{itemize}}
\newcommand{\bt}{\begin{tabular}}    \newcommand{\et}{\end{tabular}}
\newcommand{\bdm}{\begin{displaymath}} \newcommand{\edm}{\end{displaymath}}

\def\qed{\hfill{Q.E.D.}\smallskip}

\newcommand{\ls}{\setlength{\baselineskip}{12pt}
                 \setlength{\parskip}{3mm}}

\begin{document}

\title[ADM and Bondi energy-momenta]{On the relation between ADM
and Bondi energy-momenta}

\author{Xiao Zhang}
\address{Institute of Mathematics, Academy of Mathematics and
Systems Science, Chinese Academy of Sciences, Beijing 100080,
China} \email{xzhang@amss.ac.cn}

\begin{abstract}
When a spacetime takes Bondi radiating metric, and is vacuum and
asymptotically flat at spatial infinity which ensures the positive
mass theorem, we prove that the standard ADM energy-momentum is
the past limit of the Bondi energy-momentum. We also derive a
formula relating the ADM energy-momentum of any asymptotically
flat spacelike hypersurface to the Bondi energy-momentum of any
null hypersurface. The formula indicates that the Bondi mass is
always less than the ADM total energy if the system has {\it
news}.

The assumed asymptotic flatness precludes gravitational radiation.
We therefore study further the relation between the ADM total
energy and the Bondi mass when gravitational radiation emits. We
find that in this case the ADM total energy is no longer the past
limit of the Bondi mass. They differ by certain quantity relating
to the {\it news} of the system.
\end{abstract}


%\subjclass{}
%\thanks{}
%\date{November 18, 2003}

\maketitle \pagenumbering{arabic}

%%%%%%%%%%%%%%%%%%%%%%%%%%%%%%%%%%%%%%%%%%%%%%%%%%%%%%%%%%%%%%%%%%%%

\mysection{Introduction}
\ls

It is well-known that the ADM total energy and total linear
momentum can be defined on asymptotically flat spatial infinity in
a spacetime \cite{ADM}. The fundamental positive mass conjecture
in general relativity, which was proved by Schoen-Yau, and later
by Witten, asserts that the ADM energy-momentum is always timelike
for a nontrivial spacetime \cite{SY1, SY2, SY3, W}. In \cite{Z1},
the positive mass theorem was extended to the case involving the
total angular momentum. This extension actually relates to the
Einstein-Cartan theory. On the other hand, in the pioneering work
of Bondi, van der Burg, Metzner and Sachs on the gravitational
waves in vacuum spacetimes, the Bondi mass associated to each null
cone is defined and their main result asserts this Bondi mass is
always non-increasing with respect to the retarded time \cite{BBM,
S, vdB}.

One main problem in general relativity is to understand what
exactly happens on the energy-momentum when an asymptotically flat
spacelike hypersurface goes to a null hypersurface. Physically, it
is believed that gravitational radiation occurs, and the energy of
the system will be carried away by gravitational waves. When a
spacetime can be conformally compactified, and asymptotically
empty and flat at null and spatial infinity in the sense of
\cite{AH}, Ashtekar and Magnon-Ashtekar demonstrated the mass at
spatial infinity is the past limit of the Bondi mass taken as the
cut approaches the ``point'' of spatial infinity \cite{AM}. This
result was later verified, in the framework of Penrose, by Hayward
replacing the Penrose conformal factor by a product of advanced
and retarded conformal factors \cite{H}, and by Valiente Kroon
using a representation of spatial infinity based on the properties
of conformal geodesics \cite{FK, V1, V2}. In \cite{CK},
Christodoulou and Klainerman proved the global existence of
globally hyperbolic, strongly asymptotically flat, maximal
foliated vacuum Einstein equations, and proved rigorously the ADM
mass at spatial infinity is the past limit of the Bondi mass. In
this paper, we assume that spacetime takes vacuum Bondi radiating
metric. We define the spatial infinity as the  ``real'' time
slice. Under the asymptotic flatness conditions at spatial
infinity which ensures the positive mass theorem, we use a
complete and rigorous argument to demonstrate that the standard
ADM energy-momentum is the past limit of the ``standard'' Bondi
energy-momentum defined in \cite{BBM, S}. We derive a formula
relating the ADM energy-momentum for a spacelike hypersurface at
time $t _0$ to the Bondi energy-momentum for a null hypersurface
at retarded time $u _0$. As a consequence, we prove that the Bondi
mass is always less than the ADM total energy if the system has
{\it news}.

However, it is presumably believed the assumptions of asymptotic
flatness at spatial infinity in all above works precludes
gravitational radiation, at least near spatial infinity. We
therefore assume certain weaker conditions on asymptotic flatness
at spatial infinity which spacetimes may include gravitational
radiation. We also derive a formula relating the ADM total energy
to the Bondi mass. We find that, in this case, the ADM total
energy is no longer the past limit of the Bondi mass and they
differ by certain quantity relating to the {\it news} of the
system.

It should be pointed out that the ``real'' time $t$ is assumed to
be the retarded time $u$ plus the Euclidean distance $r$ in the
paper. This condition is very restricted which is not satisfied
even in Schwartzschild spacetime. In the forthcoming paper \cite{HZ},
we will study the more general case which $t$ approaches to
$u+r$ in certain sense.

The paper is organized as follows: In Section 2, we state some
well-known formulation and results of Bondi, van der Burg, Metzner
and Sachs. In Section 3, we give some asymptotically flat
conditions on Bondi radiating metric, which ensures the ADM total
energy is well-defined at spatial infinity. In Section 4, we
derive the second fundamental form of spatial infinity and show
that it is also asymptotically flat, which ensures the ADM total
linear momentum is well-defined at spatial infinity. In Section 5,
we prove that the ADM total energy is the Bondi mass of negatively
infinite retarded time. In Section 6, we prove that the ADM total
linear momentum is the Bondi momentum of negatively infinite
retarded time. In Section 7, we establish a relation between the
ADM total energy-momentum of spatial infinity at any time and the
Bondi energy-momentum at any retarded time. We also prove that the
Bondi mass is always less than the ADM total energy if the system
has {\it news}. In Section 8, we establish a relation between the
ADM total energy and the Bondi mass for Bondi radiating metric
which includes gravitational radiation.

%%%%%%%%%%%%%%%%%%%%%%%%%%%%%%%%%%%%%%%%%%%%%%%%%%%%%%%%%%%%%%%%%%%%%%%%%%%%
\mysection{The Bondi coordinates}
\ls

Throughout the paper we assume that $\big(L ^{3,1}, \tilde g\big)
$ is a vacuum spacetime with metric $\tilde g =\tilde g _{ij} dx
^i dx ^j$ taking the following Bondi radiating metric
 \beq
\tilde g &=&\Big(\frac{V}{r} e ^{2\beta} +r ^2 e ^{2 \gamma} U ^2
\cosh 2\delta +r ^2 e ^{-2 \gamma} W ^2 \cosh 2\delta
\nonumber\\
&&+2 r ^2 UW \sinh 2 \delta \Big)du ^2 -2e ^{2\beta}
du dr   \nonumber\\
& &-2r ^2 \Big(e ^{2 \gamma} U \cosh 2\delta
+W \sinh 2 \delta \Big) du d\theta     \nonumber\\
& &-2r ^2
\Big(e ^{-2 \gamma} W \cosh 2\delta+U \sinh 2\delta \Big)\sin
\theta du d\psi    \nonumber\\
& &+r ^2 \Big(e ^{2 \gamma} \cosh
2\delta d\theta ^2 +e ^{-2 \gamma}\cosh 2\delta \sin ^2 \theta d
\psi ^2   \nonumber\\
& &+2 \sinh 2\delta \sin \theta d \theta d
\psi \Big) \label{bondi-metric}
 \eeq
and satisfies the outgoing radiation condition, where $\beta,
\gamma, \delta, U, V, W$ are functions of
 \beQ
x ^0 =u,\;\;x ^1=r,\;\;x ^2=\theta, \;\;x ^3=\psi.
 \eeQ
$u$ is a retarded coordinate, $r$ is Euclidean distance, $\theta $
and $\psi $ are spherical coordinates, $0 \leq \theta \leq \pi$,
$0 \leq \psi \leq 2 \pi$. We assume the ``real'' time $t$ is
defined as
 \beQ
t=u+r.
 \eeQ
In the forthcoming paper \cite{HZ}, we will study the more general
relation between the ``real'' time $t$ and the retarded time $u$.
The metric (\ref{bondi-metric}) was studied by Bondi, van der
Burg, Metzner and Sachs in the theory of gravitational waves in
general relativity \cite{BBM, S, vdB}. They proved that the
following asymptotic behavior holds
 \beQ
\gamma &=&\frac{c(u, \theta, \psi)}{r}+O\Big(\frac{1}{r ^3}\Big),  \\
\delta &=&\frac{d(u, \theta, \psi)}{r}+O\Big(\frac{1}{r ^3}\Big),  \\
\beta &=&-\frac{c ^2 + d ^2}{4r ^2} +O\Big(\frac{1}{r ^4}\Big),     \\
U &=& -\frac{l(u, \theta, \psi)}{r ^2}+O\Big(\frac{1}{r ^3}\Big),  \\
W &=& -\frac{\bar l(u, \theta, \psi)}{r ^2}+O\Big(\frac{1}{r ^3}\Big),  \\
V &=& -r +2 M (u, \theta, \psi) +O\Big(\frac{1}{r}\Big),
 \eeQ
where
 \beQ
l &=& c _{, 2} +2c \cot \theta +d _{, 3} \csc \theta,\\
\bar l &=& d _{, 2} +2d \cot \theta -c _{,3} \csc \theta.
 \eeQ
(Throughout the paper, denote $f _{,i} = \frac{\partial
f}{\partial x ^i }$ for $i=0,1,2,3$.) $M$ is the {\it mass aspect}
and $c_{,0}$, $d _{, 0}$ are the {\it news functions} and they
satisfy the following equation \cite{vdB}:
 \beq
M _{,0} =-\Big((c _{,0} )^2 +(d _{, 0} )^2 \Big)
+\frac{1}{2}\Big(l _{,2} +l \cot \theta +\bar l _{,3} \csc \theta
\Big) _{,0} .\label{u-deriv}
 \eeq
To avoid the singularity, we assume
 \begin{description}
\item[Condition A] {\bf Each of the six functions $\beta $, $\gamma$,
$\delta$, $U$, $V$, $W$ together with its derivatives up to the
second orders are equal at $\psi =0$ and $2\pi$}.
 \end{description}
This implies
 \beQ
M \big | _{\psi =0}=M \big | _{\psi =2\pi}, && M _{,p} \big |
_{\psi =0}=M _{,p}\big | _{\psi =2\pi}, \label{sing1} \\
c \big | _{\psi =0}=c\big | _{\psi =2\pi}, && c _{,p} \big |
_{\psi =0}=c _{,p}\big | _{\psi =2\pi}, \label{sing2}\\
d \big | _{\psi =0}=d\big | _{\psi =2\pi}, && d _{,p} \big |
_{\psi =0}=d _{,p}\big | _{\psi =2\pi} \label{sing3}
 \eeQ
for $p, q=0, 2, 3$. Denote $S ^2$ the unit 2-sphere. The physical
reason requires (e.g. \cite{BBM})
 \begin{description}
\item[Condition B] {\bf For all $u$,
 \beQ
\int _0 ^{2\pi} c(u, 0, \psi) d\psi =0, \;\; \int _0 ^{2\pi} c(u,
\pi, \psi) d\psi =0. \label{sing4}
 \eeQ}
 \end{description}

Let $N _{u _0}$ be a null hypersurface which is given by $u=u _0$
at null infinity. The Bondi energy-momentum of $N _{u _0}$ is
defined by \cite{BBM, CJM}:
 \beQ
m _\nu (u _0) = \frac{1}{4 \pi} \int _{S ^2} M (u _0, \theta,
\psi) n ^{\nu} d S
 \eeQ
where $\nu =0, 1, 2, 3$, $n ^0 =1$, $n ^i $ the restriction of the
natural coordinate $x ^i$ to the unit round sphere, i.e.,
 \beQ
n ^0 =1,\;\; n ^1 = \sin \theta \cos \psi,\;\; n ^2 = \sin \theta
\sin \psi,\;\; n ^3 = \cos \theta.
 \eeQ
$m _0$ is referred as the Bondi mass. Under {\bf Condition A} and
{\bf Condition B},
 \beQ
& &\int _{S ^2}\Big(l _{,2} +l \cot \theta +\bar l _{,3} \csc
\theta \Big) dS\\
&=&\int _{0} ^{\pi} \int _{0} ^{2\pi} \Big(l _{,2} \sin
\theta +l \cos \theta +\bar l _{,3} \Big)d\psi d \theta \\
&=&\int _0 ^{2\pi} \Big(l \sin \theta \Big)\Big | _{\theta =0}
^{\pi} d \psi +\int _{0} ^{\pi} \Big(\bar l (u, \theta, 2\pi)
-\bar l (u, \theta, 0)\Big)d \theta\\
&=&-2 \int _0 ^{2\pi} \Big(c (u, \pi, \psi)+c(u, 0, \psi)\Big)d
\psi\\
&=&0,
 \eeQ
then (\ref{u-deriv}) gives rise to the famous Bondi mass loss
formula
 \beq
\frac{d}{du} m _0 =-\frac{1}{4 \pi} \int _{S ^2} \Big((c _{,0} )^2
+(d _{,0} )^2 \Big) d S.  \label{u-Bondi-mass}
 \eeq
Now we derive the Bondi momentum loss formula. It is easy to find
 \beQ
 & &\int _{S ^2}\Big(l _{,2} +l \cot \theta +\bar l _{,3}
\csc
\theta \Big) n ^1 dS \\
&=&\int _{0} ^{\pi} \int _{0} ^{2\pi} \Big(l _{,2}
\sin\theta +l \cos \theta +\bar l _{,3} \Big)\sin \theta \cos \psi
d\psi d \theta \\
&=&\int _0 ^{2\pi} \Big( \big(l \sin ^2 \theta \big) \Big|
_{\theta =0} ^{\pi} -\int _0 ^{\pi} l \sin \theta \cos \theta d
\theta \Big)\cos \psi d \psi\\
& &+\int _0 ^{\pi} \Big( \big(\bar l \cos \psi \big) \Big| _{\psi
=0} ^{2\pi} +\int _0 ^{2\pi} \bar l \sin \psi d
\psi \Big)\sin \theta d \theta\\
&=&-\int _{0} ^{\pi} \int _{0} ^{2\pi} \Big(c\cos\psi+d \cos\theta
\sin \psi \Big)d\psi d \theta\\
& &+\int _{0} ^{\pi} \int _{0} ^{2\pi} \Big(c\cos\psi+d \cos\theta
\sin \psi \Big)d\psi d \theta\\
&=&0,
 \eeQ
and
 \beQ
 & &\int _{S ^2}\Big(l _{,2} +l \cot \theta +\bar l _{,3}
\csc
\theta \Big) n ^2 dS \\
&=&\int _{0} ^{\pi} \int _{0} ^{2\pi} \Big(l _{,2}
\sin\theta +l \cos \theta +\bar l _{,3} \Big)\sin \theta \sin \psi
d\psi d \theta \\
&=&\int _0 ^{2\pi} \Big( \big(l \sin ^2 \theta \big) \Big|
_{\theta =0} ^{\pi} -\int _0 ^{\pi} l \sin \theta \cos \theta d
\theta \Big)\sin \psi d \psi\\
& &+\int _0 ^{\pi} \Big( \big(\bar l \sin \psi \big) \Big| _{\psi
=0} ^{2\pi} -\int _0 ^{2\pi} \bar l \cos \psi d
\psi \Big)\sin \theta d \theta\\
&=&-\int _{0} ^{\pi} \int _{0} ^{2\pi} \Big(c\sin\psi-d \cos\theta
\cos \psi \Big)d\psi d \theta\\
& &+\int _{0} ^{\pi} \int _{0} ^{2\pi} \Big(c\sin\psi-d \cos\theta
\cos \psi \Big)d\psi d \theta\\
&=&0,
 \eeQ
and
 \beQ
 & &\int _{S ^2}\Big(l _{,2} +l \cot \theta +\bar l _{,3}
\csc
\theta \Big) n ^3 dS \\
&=&\int _{0} ^{\pi} \int _{0} ^{2\pi} \Big(l _{,2}
\sin\theta +l \cos \theta +\bar l _{,3} \Big)\cos \theta d\psi d \theta \\
&=&\int _{0} ^{\pi} \int _{0} ^{2\pi} \Big(l \sin ^2 \theta
\Big)d\psi d \theta + \int _{0} ^{2\pi} \Big(l \sin \theta \cos
\theta \Big)
\Big | _{\theta =0} ^{\pi} d\psi\\
&=& 2\int _{0} ^{2\pi} \Big(c(u, \pi, \psi)-c(u, 0, \psi) \Big)
d\psi\\
&=&0.
 \eeQ
We obtain, for $k$=$1$, $2$, $3$,
 \beq
\frac{d}{du} m _k =-\frac{1}{4 \pi} \int _{S ^2} \Big((c _{,0}) ^2
+(d _{,0}) ^2 \Big) n ^k d S. \label{u-Bondi-momentum}
 \eeq

In general, the spatial infinity in vacuum spacetimes which the
metric satisfies (\ref{bondi-metric}) may not be asymptotically
flat in the sense of \cite{SY1, SY2, SY3, W, PT, Z1}. Using $(t,
r, \theta, \psi)$ coordinates, the metric (\ref{bondi-metric}) can
be written as
 \beq
\tilde g &=&\Big(\frac{V}{r} e ^{2\beta} +r ^2 e ^{2 \gamma} U
^2 \cosh 2\delta    \nonumber\\
& &+r ^2 e ^{-2 \gamma} W ^2 \cosh 2\delta +2 r ^2 UW \sinh 2
\delta\Big)dt ^2   \nonumber\\
& & -2\Big(\big(1+\frac{V}{r}\big) e ^{2\beta} +r ^2 e ^{2 \gamma}
U ^2 \cosh 2\delta    \nonumber\\
& &+r ^2 e ^{-2 \gamma} W ^2 \cosh 2\delta +2 r ^2 UW \sinh 2
\delta \Big) dt dr    \nonumber\\
& &-2 r ^2 \Big(e ^{2 \gamma } U \cosh 2 \delta +W \sinh 2 \delta
\Big)dt d\theta
\nonumber\\
& &-2 r ^2 \Big(e ^{-2 \gamma } W \cosh 2 \delta +U \sinh 2 \delta
\Big)\sin \theta dt d\psi
\nonumber\\
 & & +\Big(\big(2+\frac{V}{r}\big) e ^{2\beta} +r ^2 e ^{2 \gamma}
U ^2 \cosh 2\delta      \nonumber\\
& & +r ^2 e ^{-2 \gamma} W ^2 \cosh 2\delta +2 r ^2 UW \sinh 2
\delta \Big) dr ^2
 \nonumber\\
 & &+r ^2 \Big(e ^{2 \gamma} \cosh 2\delta d\theta ^2 +e
^{-2\gamma}\cosh 2\delta \sin ^2 \theta d \psi ^2   \nonumber\\
& &+2 \sinh 2\delta \sin \theta d \theta d \psi \Big)
 \nonumber\\
& &+2 r ^2 \Big(e ^{2 \gamma } U \cosh 2 \delta +W \sinh 2 \delta
\Big)dr d\theta     \nonumber\\
& &+2 r ^2 \Big(e ^{-2 \gamma } W \cosh 2 \delta +U \sinh 2 \delta
\Big)\sin \theta dr d\psi . \label{spacetime-metric}
 \eeq
%%%%%%%%%%%%%%%%%%%%%%%%%%%%%%%%%%%%%%%%%%%%%%%%%%%%%%%%%%%%%%%%%%%%%%%%%%%%
\mysection{The spatial infinity}
\ls

Let $\big(N _{t _0}, g, h\big)$ be a spacelike hypersurface in $L
^{3,1}$ which is given by $\big\{t= t _0\big\}$, where $g$ is the
induced metric of $\tilde g$ and $h$ is the second fundamental
form.
 \beq
g&=&\Big(\big(2+\frac{V}{r}\big) e ^{2\beta} +r ^2 e ^{2 \gamma} U
^2 \cosh 2\delta    \nonumber\\
& &+r ^2 e ^{-2 \gamma} W ^2 \cosh 2\delta +2 r ^2 UW \sinh 2
\delta  \Big) dr ^2   \nonumber\\
& & +r ^2 \Big(e ^{2 \gamma} \cosh 2\delta d\theta ^2
 +e ^{-2\gamma}\cosh 2\delta \sin ^2 \theta d \psi ^2 \nonumber\\
 & & +2 \sinh 2\delta \sin \theta d \theta d \psi \Big)
 \nonumber\\
& &+2 r ^2 \Big(e ^{2 \gamma } U \cosh 2 \delta +W \sinh 2 \delta
\Big)dr d\theta     \nonumber\\
& &+2 r ^2 \Big(e ^{-2 \gamma } W \cosh 2 \delta +U \sinh 2 \delta
\Big)\sin \theta dr d\psi . \label{metric}
 \eeq
$\big(N _{t_0}, g, h\big)$ is usually refereed to an initial data
set. We will study when $\big(N _{t _0}, g, h\big)$ is
asymptotically flat. Let $\{ \breve{e} ^i \}$ ($i=1,2,3$) be the
coframe of the standard flat metric $g _0$ on $\R ^3$,
 \beQ
\breve{e} ^1 = dr, \;\;\breve{e} ^2 = r d\theta, \;\;\breve{e} ^3
= r \sin \theta d\psi.
 \eeQ
Let $\{\breve{e} _i \}$ ($i=1,2,3$) be the dual frame. The
connection 1-form $\{\breve{\omega } _{ij}\}$ is given by $d
\breve{e} ^i= - \breve{\omega} _{ij} \wedge \breve{e} ^{j}$, or
$\breve{\nabla} \breve{e} _{i}= - \breve{\omega} _{ij} \otimes
\breve{e} _{j}$ ($i,j=1,2,3$) where $\breve{\nabla}$ is
Levi-Civita connection of $g _0$. It is easy to find that
 \beQ
    \breve{\omega}  _{12} =  -\frac{1}{r}\breve{e}  ^{2}, \;\;
    \breve{\omega}  _{13} =  -\frac{1}{r}\breve{e}  ^{3}, \;\;
    \breve{\omega}  _{23} =  -\frac{\cot \theta}{r}\breve{e} ^{3}.
 \eeQ
Throughout the paper, we denote $\breve{\nabla} _i \equiv
\breve{\nabla} _{\breve{e} _i}$ for $i=1,2,3$.

The initial data set $\big(N _{t _0}, g, h \big)$ is
asymptotically flat in the current case if the metric $g$
satisfies
 \beq
 g\big(\breve{e} _i, \breve{e} _j\big)=\delta
 _{ij}+O\Big(\frac{1}{r}\Big),
 \breve{\nabla} _k g\big(\breve{e} _i, \breve{e} _j\big)
 =O\Big(\frac{1}{r^2}\Big),
 \breve{\nabla} _l \breve{\nabla} _k  g\big(\breve{e} _i, \breve{e} _j\big)
 =O\Big(\frac{1}{r^3}\Big)
 \label{gij}
 \eeq
as $r \rightarrow \infty$. Furthermore, $2$-tensor $h$ satisfies
 \beq
 h \big(\breve{e} _i, \breve{e} _j\big) =O\Big(\frac{1}{r^2}\Big),
 \breve{\nabla} _k  h \big(\breve{e} _i, \breve{e} _j\big) =
 O\Big(\frac{1}{r^3}\Big) \label{hij}
 \eeq
as $r \rightarrow \infty$.

Denote $\mathcal{C} _{\{a_1, a_2, a_3\}}$ the functions in
spacetime which satisfy the following asymptotic behavior at
spatial infinity
 \beq
\mathcal{C} _{\{a_1, a_2, a_3\}} =
 \left\{f:
 \begin{array}{ccc}
    \lim _{r \rightarrow \infty}\lim _{u \rightarrow -\infty} r ^{a_1}f
    &=&O\big(1\big),\\
    \lim _{r \rightarrow \infty}\lim _{u \rightarrow -\infty} r ^{a_2}
    \breve{\nabla} _i f
    &=&O\big(1\big),\\
    \lim _{r \rightarrow \infty}\lim _{u \rightarrow -\infty} r ^{a_3}
    \breve{\nabla} _i\breve{\nabla} _j f &=& O\big(1\big)
 \end{array}
 \right\}.       \label{mathcal-C}
 \eeq

We employ the following assumptions:
 \begin{description}
\item[Condition C]
$\;\;\gamma \in \mathcal{C} _{\{1, 2, 3\}},\;\; \delta \in
\mathcal{C} _{\{1, 2, 3\}}, \;\;\beta \in \mathcal{C} _{\{2, 3,
4\}}, \;\;U \in \mathcal{C} _{\{2, 3, 4\}}, \;\;W \in \mathcal{C}
_{\{2, 3, 4\}}, \;\;V+r \in \mathcal{C} _{\{0, 1, 2\}}$.
 \end{description}
{\bf Condition C} implies, for $r$ sufficiently large,
 \beQ
& &\lim _{u \rightarrow -\infty}M=O\Big(1\Big),
\lim _{u \rightarrow -\infty}c=O\Big(1\Big),
\lim _{u \rightarrow -\infty}d=O\Big(1\Big),\\
& &\lim _{u \rightarrow -\infty}M _{,0}=O\Big(\frac{1}{r}\Big),
\lim _{u \rightarrow -\infty}c _{,0}=O\Big(\frac{1}{r}\Big),
\lim _{u \rightarrow -\infty}d _{,0}=O\Big(\frac{1}{r}\Big),\\
& &\lim _{u \rightarrow -\infty}M _{,A}=O\Big(1\Big), \lim _{u
\rightarrow -\infty}c _{,A}=O\Big(1\Big), \lim _{u \rightarrow
-\infty}d _{,A}=O\Big(1\Big).
 \eeQ
where $A,B=2, 3$.
 \begin{prop}\label{g}
Under {\bf Condition A}, {\bf Condition B} and {\bf Condition C},
the metric $g$ of $N _{t _0}$ satisfies (\ref{gij}).
 \end{prop}
\pf The components of the metric $g$ are
 \beQ
g\big(\breve{e} _1, \breve{e}
_1\big)&=&\Big[\big(2+\frac{V}{r}\big) e
^{2\beta} +r ^2 e ^{2 \gamma} U ^2 \cosh 2\delta\\
& &+r ^2 e ^{-2 \gamma} W ^2 \cosh 2\delta +2 r ^2 UW \sinh 2
\delta \Big] _{t=t _0} \\
g\big(\breve{e} _2, \breve{e} _2\big)&=& e ^{2 \gamma} \cosh
2\delta \Big| _{t=t _0}\\
g\big(\breve{e} _3, \breve{e} _3\big)&=& e ^{-2\gamma} \cosh 2\delta \Big| _{t=t _0}\\
g\big(\breve{e} _1, \breve{e} _2\big)&=& r \Big(e ^{2 \gamma } U
\cosh 2 \delta +W \sinh 2 \delta\Big) _{t=t _0}\\
g\big(\breve{e} _1, \breve{e} _3\big)&=& r \Big(e ^{-2 \gamma } W
\cosh 2 \delta +U \sinh 2 \delta\Big) _{t=t _0}\\
g\big(\breve{e} _2, \breve{e} _3\big)&=& \sinh 2\delta \Big| _{t=t
_0}.
 \eeQ
Note that for fixed $t=t _0$, $r \rightarrow \infty$ is equivalent
to $u \rightarrow -\infty$, a straightforward computation yields
the proposition.   \qed
%%%%%%%%%%%%%%%%%%%%%%%%%%%%%%%%%%%%%%%%%%%%%%%%%%%%%%%%%%%%%%%%%%%%%%%%%%%%
\mysection{The second fundamental form}
\ls

The lapse $\N$ and the shift $X _i$ $(i=1,2,3)$ of the spacelike
hypersurface $N _{t _0}$ are
 \beQ
\N &=&\Big(-\tilde g ^{tt} \Big) ^{-\frac{1}{2}} _{t=t_0}, \\
X _i &=&\tilde g _{ti} \Big| _{t=t_0}.
 \eeQ
The second fundamental form is then given by
 \beQ
h _{ij} =\frac{1}{2 \N} \Big(\nabla _i X _j +\nabla _j X _i
-\partial _t \tilde g _{ij} \Big)
 \eeQ
where
 \beQ
\nabla _i X _j =\partial _i X _j -\Gamma ^k _{ij} X _k
 \eeQ
and
 \beQ
\Gamma _{ij} ^k =\frac{1}{2} g ^{kl} \Big(\frac{\partial g
_{li}}{\partial x ^j} +\frac{\partial g _{lj}}{\partial x ^i}
-\frac{\partial g _{ij}}{\partial x ^l} \Big)_{t=t_0}
 \eeQ
are Christoffel symbols of the metric $g$. Now we compute the
inverse $g ^{ij}$ of metric tensor $g _{ij}$ of $N _{t_0}$. Denote
$\bar g =\Big( g _{AB} \Big)$, $2 \leq A, B \leq 3$, i.e.,
 \beQ
\bar g = r ^2 \left(\begin{array}{cc}
  e ^{2 \gamma }\cosh 2\delta &  \sinh 2\delta \sin \theta\\
  \sinh 2\delta \sin \theta  &  e ^{-2 \gamma }\cosh 2\delta \sin
  ^2 \theta
              \end{array}  \right).
 \eeQ
Then the inverse $\bar g ^{-1} =\Big( g ^{AB} \Big)$
 \beQ
\bar g ^{-1}=\frac{1}{r ^2}\left(\begin{array}{cc}
 e ^{-2 \gamma }\cosh 2\delta & -\frac{\sinh 2\delta}{\sin \theta} \\
 -\frac{\sinh 2\delta }{\sin \theta } & e ^{2 \gamma }\frac{\cosh
 2\delta}{\sin ^2 \theta}
              \end{array}  \right).
 \eeQ
Using the formulae (e.g. \cite{CJK})
 \beQ
\frac{1}{g ^{11}}&=&g _{11}-\bar g ^{AB} g _{1B} g _{1A},\\
\frac{g ^{1A}}{g ^{11}}&=&-\bar g ^{AB} g _{1B} ,\\
g ^{AB}&=&\bar g ^{AB} +\frac{g _{1A} g _{1B}}{g ^{11}},
 \eeQ
we obtain
 \beQ
\frac{1}{g ^{11}}&=&\big(2+\frac{V}{r}\big) e ^{2\beta},\\
g ^{12} &=& -U g ^{11},\\
g ^{13} &=& -\frac{W}{\sin \theta} g ^{11},\\
g ^{22} &=& \frac{e ^{-2 \gamma} \cosh 2\delta}{r ^2} +U ^2 g^{11},\\
g ^{23} &=& -\frac{ \sinh 2\delta}{r ^2 \sin \theta} +\frac{UW}{\sin \theta} g^{11},\\
g ^{33} &=& \frac{e ^{2 \gamma} \cosh 2\delta}{r ^2 \sin ^2
\theta} +\frac{W ^2}{\sin ^2 \theta} g^{11}.
 \eeQ
 \begin{prop}\label{h}
Under {\bf Condition A}, {\bf Condition B} and {\bf Condition C},
the second fundamental form $h$ of $N _{t _0}$ satisfies
(\ref{hij}).
 \end{prop}
\pf With the help of asymptotic behavior of $\beta, \gamma, \delta, U,
V, W$, we obtain the asymptotic expansion of the Christoffel
symbols of (\ref{metric}) (see Appendix). And a straightforward
computation yields
 \beQ
 \N ^2&=&-\tilde g _{tt} +\tilde g _{ti} \tilde g _{tj} g ^{ij}\\
      &=&e ^{4 \beta} g ^{11}\\
      &=&1-\frac{2M}{r} \Big|_{t=t_0}+O\Big(\frac{1}{r ^2}\Big).\\
X _1 &=& -\Big(1+\frac{V}{r}\Big) e ^{2 \beta} -r ^2 e ^{2 \gamma}
U ^2 \cosh 2\delta \\
& &-r ^2 e ^{-2 \gamma} W ^2 \cosh 2\delta -2 r ^2 UW \sinh 2 \delta \\
&=&-\frac{2M}{r}\Big|_{t=t_0}+O\Big(\frac{1}{r ^2}\Big),\\
X _2&= &- r ^2 \Big(e ^{2 \gamma } U \cosh 2 \delta +W \sinh 2
\delta
\Big)\\
 &=&l \Big|_{t=t_0}+O\Big(\frac{1}{r}\Big),\\
X _3&= &- r ^2 \Big(e ^{-2 \gamma } W \cosh 2 \delta +U \sinh 2
\delta
\Big)\sin \theta \\
&=&\bar l \sin \theta \Big|_{t=t_0}+O\Big(\frac{1}{r}\Big).
 \eeQ
We obtain the second fundamental forms
 \beQ
h _{11}&=&\frac{2M+r M _{,0}}{r ^2}\Big| _{t=t_0}+O\Big(\frac{1}{r ^3}\Big),\\
h _{22}&=&\Big(l _{, 2} -2 M -r c _{, 0}\Big) _{t=t_0}+O\Big(\frac{1}{r}\Big),\\
h _{33}&=&\Big(\bar l _{, 3} \sin \theta-2M \sin ^2 \theta +l
\sin \theta \cos \theta
        +r c _{, 0} \sin ^2 \theta \Big) _{t=t_0}+O\Big(\frac{1}{r}\Big),\\
h _{12}&=&-\frac{M _{,2} +l}{r}\Big| _{t=t_0}+O\Big(\frac{1}{r ^2}\Big),\\
h _{13}&=&-\frac{M _{,3} +l \sin \theta }{r}\Big| _{t=t_0}+O\Big(\frac{1}{r ^2}\Big),\\
h _{23}&=&\Big(\frac{\bar l _{,2} \sin \theta -\bar l \cos \theta
+l _{,3}}{2}-r d _{,0} \sin \theta \Big) _{t=t_0}
+O\Big(\frac{1}{r}\Big).
 \eeQ
%
%
% \beQ
%h\big(\breve{e} _1, \breve{e} _1\big)
%&=&\frac{2M+r M _{,0}}{r ^2}\Big| _{t=t_0}+O\Big(\frac{1}{r ^3}\Big),\\
%h\big(\breve{e} _2, \breve{e} _2\big)
%&=&\frac{l _{, 2} -2 M -r c _{, 0}}{r ^2}\Big| _{t=t_0}+O\Big(\frac{1}{r ^3}\Big),\\
%h\big(\breve{e} _3, \breve{e} _3\big)
%&=&\frac{\bar l _{, 3} \sin
%\theta-2M \sin ^2 \theta +l \sin \theta \cos \theta
%        +r c _{, 0} \sin ^2 \theta }{r^2 \sin ^2 \theta}\Big| _{t=t_0}
%        +O\Big(\frac{1}{r^3}\Big),\\
%h\big(\breve{e} _1, \breve{e} _2\big)
%&=&-\frac{M _{,2} +l}{r ^2}\Big| _{t=t_0}+O\Big(\frac{1}{r ^3}\Big),\\
%h\big(\breve{e} _1, \breve{e} _3\big) &=&-\frac{M _{,3} +l \sin
%\theta }{r ^2 \sin \theta}\Big| _{t=t_0}
%+O\Big(\frac{1}{r ^3}\Big),\\
%h\big(\breve{e} _2, \breve{e} _3\big)&=&\frac{\bar l _{,2} \sin
%\theta -\bar l \cos \theta +l _{,3}-2r d _{,0} \sin \theta}{2r ^2
%\sin \theta}\Big| _{t=t_0}+O\Big(\frac{1}{r ^3}\Big).
% \eeQ
%
%
Note that for fixed $t=t _0$, $r \rightarrow \infty$ is equivalent
to $u \rightarrow -\infty$, a straightforward computation yields
the proposition. \qed

The trace of the second fundamental form is
 \beQ
tr _g \big(h\big)=\frac{1}{r ^2} \Big(r M _{,0} -2M +l \cot \theta
+l _{,2 } +\bar l _{, 3} \csc \theta \Big)
_{t=t_0}+O\Big(\frac{1}{r ^3}\Big).
 \eeQ
%%%%%%%%%%%%%%%%%%%%%%%%%%%%%%%%%%%%%%%%%%%%%%%%%%%%%%%%%%%%%%%%%%%%%%%%%%%%
\mysection{The ADM total energy} \ls

In this section, we compute the ADM total energy of $N _{t _0}$.
In polar coordinates, the ADM total energy $\mathbb{E}$ is
 \beQ
\mathbb{E}=\frac{1}{16\pi}\lim _{r \rightarrow \infty} \int _{S
_r} \Big(\breve {\nabla } ^j g\big(\breve{e} _1, \breve{e} _j\big)
-\breve {\nabla } _1 tr _{g _0} \big(g\big) \Big) \breve{e} ^2
\wedge \breve{e} ^3.
 \eeQ
This can be seen by changing Euclidean coordinates to the polar
coordinates, or by writing Witten's mass formula \cite{W, PT, Z1}
in asymptotically polar coordinates and using the comparison of
two spin connections (e.g., \cite{Z3}). The polar coordinate
expression of the ADM total energy is equivalent to the Euclidean
coordinate expression of \cite{ADM} even if the metric is not the
standard asymptotically flat in the sense of \cite{SY1, SY2, SY3,
W, PT, Z1}. This is because the coordinate transformation relates
to only the ground metric which is the standard metric of $\R ^3$.
 \begin{thm}\label{energy}
Under {\bf Condition A}, {\bf Condition B} and {\bf Condition C},
the ADM total energy of $N _{t _0}$ is
 \beQ
\mathbb{E} (t_0) = m _0 (-\infty).
 \eeQ
 \end{thm}
\pf By (\ref{metric}), we obtain
 \beQ
\breve {\nabla } ^j g\big(\breve{e} _1, \breve{e} _j\big) -\breve
{\nabla } _1 tr _{g _0} \big(g\big)
 &=&\breve {e} _j \Big(g\big(\breve{e} _1, \breve{e} _j\big)\Big)
 -\breve{e} _1 tr _{g _0} \big(g\big)\\
 & &-g\big(\breve{e} _j, \breve{e} _i\big)\breve {\omega} _{i1}
 \big(\breve{e} _j \big)\\
 & &-g\big(\breve{e} _1, \breve{e} _i\big)\breve {\omega} _{ij}
 \big(\breve{e} _j \big)\\
 &=&\frac{4M}{r ^2} -\frac{l _{,2}}{r ^2} -\frac{l \cot \theta }{r
 ^2}-\frac{\bar l _{,3}}{r ^2 \sin \theta}\\
 & & +O\Big(\frac{1}{r ^3}\Big).
 \eeQ
Thus
 \beQ
\mathbb{E}(t_0)&=&\frac{1}{4\pi}\lim _{r \rightarrow \infty} \int
_{S ^2} M \Big| _{t=t_0}dS\\
  & &  -\frac{1}{16\pi}\lim _{r \rightarrow \infty} \int _0 ^{\pi}
    \int _0 ^{2\pi} \Big( l _{,2} \sin \theta+l \cos \theta
    +\bar l _{,3} \Big) _{t=t_0} d\psi d\theta\\
 &=&\frac{1}{4\pi}\lim _{u \rightarrow -\infty} \int _{S ^2} M
 dS\\
 & &   -\frac{1}{16\pi}\lim _{u \rightarrow -\infty} \int _0 ^{\pi}
    \Big(\bar l (u, \theta, 2\pi) -\bar l(u, \theta, 0 )\Big)
    d\theta\\
 & &  -\frac{1}{16\pi}\lim _{u \rightarrow -\infty} \int _0 ^{2\pi}
    \Big( l \sin \theta \Big) \Big |_{\theta =0} ^{\pi}d\psi.\\
 &=&\frac{1}{4\pi} \lim _{u \rightarrow -\infty} \int _{S ^2} M(u,
\theta, \psi) dS \\
& &+\frac{1}{8\pi} \lim _{u \rightarrow -\infty} \int _0 ^{2\pi}
\Big(c(u, 0, \psi)+c(u, \pi, \psi) \Big)d \psi\\
&= &m _0 (-\infty).
 \eeQ   \qed
%%%%%%%%%%%%%%%%%%%%%%%%%%%%%%%%%%%%%%%%%%%%%%%%%%%%%%%%%%%%%%%%%%%%%%%
\mysection{The ADM total linear momentum}
\ls

In this section, we compute the ADM total linear momentum of $N _{t
_0}$. Let Euclidean coordinates
 \beQ
y ^1 =r \sin \theta \cos \psi,\;\;\;\; y ^2=r \sin \theta \sin
\psi, \;\;\;\;y ^3 =r \cos \theta.
 \eeQ
Then the ADM total linear momentum
 \beQ
\mathbb{P} _i =\frac{1}{8\pi}\lim _{r \rightarrow \infty} \int _{S
_r} \Big(h\Big(\frac{\partial}{\partial y ^i},
\frac{\partial}{\partial r}\Big) -g\Big(\frac{\partial}{\partial y
^i}, \frac{\partial}{\partial r}\Big)tr _{g} \big(h\big) \Big)
\breve{e} ^2 \wedge \breve{e} ^3.
 \eeQ
A simple computation yields
 \beQ
\frac{\partial}{\partial y ^1} &=& \frac{\partial}{\partial r} n
^1+\frac{\partial}{\partial \theta} \frac{\cos
\theta \cos \psi }{r}- \frac{\partial}{\partial \psi}
\frac{\sin \psi}{r \sin \theta},\\
\frac{\partial}{\partial y ^2} &=& \frac{\partial}{\partial r} n
^2+\frac{\partial}{\partial \theta} \frac{\cos \theta \sin \psi
}{r} +\frac{\partial}{\partial \psi} \frac{\cos \psi}{r \sin
\theta},\\
\frac{\partial}{\partial y ^3} &=& \frac{\partial}{\partial r} n
^3-\frac{\partial}{\partial \theta}\frac{\sin \theta}{r}.
 \eeQ
Therefore, under {\bf Condition A}, {\bf Condition B} and {\bf Condition C}
 \beQ
h\Big(\frac{\partial}{\partial y ^1}, \frac{\partial}{\partial r}
\Big)&=&h _{11} n ^1 +h _{21} \frac{\cos \theta \cos \psi}{r}-h
_{31} \frac{\sin \psi}{r \sin \theta}\\
&=&\frac{1}{r ^2} \Big(\big(2M+r M _{,0}\big)n ^1 -\big(M _{,2}
+l\big) \cos \theta \cos\psi\\
&& +\frac{\sin \psi}{\sin \theta } \big(M _{,3} +\bar l \sin
\theta \big)\Big)+O\Big(\frac{1}{r
^3}\Big),\\
h\Big(\frac{\partial}{\partial y ^2}, \frac{\partial}{\partial r}
\Big)&=&h _{11} n ^2 +h _{21} \frac{\cos \theta \sin \psi}{r}+h
_{31} \frac{\cos \psi}{r \sin \theta}\\
&=&\frac{1}{r ^2} \Big(\big(2M+r M _{,0}\big)n ^2 -\big(M _{,2}
+l\big) \cos \theta \sin\psi \\
&&-\frac{\cos \psi}{\sin \theta }\big(M _{,3} +\bar l \sin \theta
\big)\Big)+O\Big(\frac{1}{r
^3}\Big),\\
h\Big(\frac{\partial}{\partial y ^3}, \frac{\partial}{\partial r}
\Big)&=&h _{11} n ^3 -h _{21} \frac{\sin \theta}{r}\\
&=&\frac{1}{r ^2} \Big(\big(2M+r M _{,0}\big)n ^3 +\big(M _{,2}
+l\big) \sin \theta \Big)+O\Big(\frac{1}{r^3}\Big).\\
g\Big(\frac{\partial}{\partial y ^1}, \frac{\partial}{\partial r}
\Big)&=&g _{11} n ^1 +g _{21} \frac{\cos \theta \cos \psi}{r}-g
_{31} \frac{\sin \psi}{r \sin \theta}\\
&=&\Big(1+ \frac{2M}{r}\Big) n ^1 -l \frac{\cos \theta
\cos\psi}{r} +\bar l \frac{\sin \psi}{r}+O\Big(\frac{1}{r
^2}\Big),\\
g\Big(\frac{\partial}{\partial y ^2}, \frac{\partial}{\partial r}
\Big)&=&g _{11} n ^2 +g _{21} \frac{\cos \theta \sin \psi}{r}+g
_{31} \frac{\cos \psi}{r \sin \theta}\\
&=&\Big(1+ \frac{2M}{r}\Big)n ^2 -l \frac{\cos \theta \sin
\psi}{r}-\bar l \frac{\cos \psi}{r}+O\Big(\frac{1}{r^2}\Big),\\
g\Big(\frac{\partial}{\partial y ^3}, \frac{\partial}{\partial r}
\Big)&=&g _{11} n ^3 -g _{21} \frac{\sin \theta}{r}\\
&=&\Big(1+ \frac{2M}{r}\Big)n ^3 +l\frac{\sin \theta}{r}+
O\Big(\frac{1}{r^2}\Big).
 \eeQ
 \begin{thm}\label{momentum}
Under {\bf Condition A}, {\bf Condition B} and {\bf Condition C},
the ADM total linear momentum of $N _{t _0}$ is
 \beQ
\mathbb{P} _k (t_0)&=& m _k (-\infty)
 \eeQ
for $k$=$1$, $2$, $3$.
 \end{thm}
\pf It is straightforward that
 \beQ
\mathbb{P} _1 (t_0)&=&\frac{1}{2\pi}\lim _{u \rightarrow
-\infty}\int _{S ^2} M n ^1 dS \\ & &-\frac{1}{8\pi} \lim _{u
\rightarrow -\infty} \int _0 ^{\pi} \int _0 ^{2\pi} \Big(M
_{,2}\sin \theta \cos \theta \cos \psi -M _{,3}\sin \psi\Big)d
\psi d \theta \\ & &-\frac{1}{8\pi} \lim _{u \rightarrow -\infty}
\int _0 ^{\pi} \int _0 ^{2\pi}\Big(l _{,2}\sin ^2\theta \cos \psi
+\bar l _{, 3}\sin \theta \cos \psi \Big)d \psi d \theta \\
 & &-\frac{1}{8\pi} \lim _{u \rightarrow -\infty} \int _0 ^{\pi}
\int _0 ^{2\pi} \Big(2 l\sin \theta \cos \theta \cos \psi -\bar l
\sin\theta \sin \psi \Big)d \psi d \theta \\
&=&\frac{1}{4\pi}\lim _{u \rightarrow -\infty}\int _{S ^2} M n ^1
dS -\frac{1}{8\pi} \lim _{u \rightarrow -\infty} \int _0 ^{\pi}
\Big( \bar l \sin \theta \cos \psi\Big) \Big|_{\psi =0} ^{2\pi} d
\theta\\
& & -\frac{1}{8\pi} \lim _{u \rightarrow -\infty} \int _0 ^{2\pi}
\Big(l \sin ^2 \theta \cos \psi\Big) \Big|_{\theta =0} ^{\pi} d
\psi\\
&= &\frac{1}{4\pi}\lim _{u \rightarrow -\infty}\int _{S ^2} M n ^1
dS\\
&=&m _1 (-\infty),
 \eeQ
and
 \beQ
\mathbb{P} _2 (t_0)&=&\frac{1}{2\pi}\lim _{u \rightarrow
-\infty}\int _{S ^2} M n ^2 dS \\ & &-\frac{1}{8\pi} \lim _{u
\rightarrow -\infty} \int _0 ^{\pi} \int _0 ^{2\pi} \Big(M
_{,2}\sin \theta \cos \theta \sin \psi +M _{,3}\cos \psi \Big)d
\psi d \theta \\ & &-\frac{1}{8\pi} \lim _{u \rightarrow -\infty}
\int _0 ^{\pi} \int _0 ^{2\pi}\Big(l _{,2}\sin ^2\theta \sin \psi
+\bar l _{,3}\sin \theta \sin \psi \Big)d \psi d \theta \\
 & &-\frac{1}{8\pi} \lim _{u \rightarrow -\infty} \int _0 ^{\pi}
\int _0 ^{2\pi} \Big(2 l \sin \theta \cos \theta \sin \psi +\bar l
\sin \theta \cos \psi \Big)d \psi d \theta \\
&=&\frac{1}{4\pi}\lim _{u \rightarrow -\infty}\int _{S ^2} M n ^2
dS-\frac{1}{8\pi} \lim _{u \rightarrow -\infty} \int _0 ^{\pi}
\Big( \bar l \sin \theta \sin \psi\Big) \Big|_{\psi =0} ^{2\pi} d
\theta\\
& &-\frac{1}{8\pi} \lim _{u \rightarrow -\infty} \int _0 ^{2\pi}
\Big(l \sin ^2 \theta \sin \psi\Big) \Big|_{\theta =0} ^{\pi} d
\psi\\
&= &\frac{1}{4\pi}\lim _{u \rightarrow -\infty}\int _{S ^2} M n ^2
dS\\
& =&m _2 (-\infty),
 \eeQ
and
 \beQ
\mathbb{P} _3 (t_0)&=&\frac{1}{2\pi}\lim _{u \rightarrow
-\infty}\int _{S ^2} M n ^3 dS \\ & &+\frac{1}{8\pi} \lim _{u
\rightarrow -\infty} \int _0 ^{\pi} \int _0 ^{2\pi} \Big(M
_{,2}\sin ^2 \theta \Big)d \psi d \theta \\ & &-\frac{1}{8\pi}
\lim _{u \rightarrow -\infty} \int _0 ^{\pi} \int _0 ^{2\pi}\Big(l
_{,2}\sin \theta \cos \theta +\bar l _{,3}\cos \theta \Big)d \psi
d \theta \\
 & &+\frac{1}{8\pi} \lim _{u \rightarrow -\infty} \int _0 ^{\pi}
\int _0 ^{2\pi} \Big( l \sin ^2\theta - l \cos ^2 \theta \Big)
d \psi d \theta \\
&=&\frac{1}{4\pi}\lim _{u \rightarrow -\infty}\int _{S ^2} M n ^3
dS-\frac{1}{8\pi} \lim _{u \rightarrow -\infty} \int _0 ^{2\pi}
\Big(l \sin \theta \cos \theta \Big) \Big|_{\theta =0} ^{\pi} d
\psi\\
&= &\frac{1}{4\pi}\lim _{u \rightarrow -\infty}\int _{S ^2} M n ^3
dS\\
& &-\frac{1}{4\pi}\lim _{u \rightarrow -\infty}\int _0 ^{2 \pi}
\Big(c (u, \pi, \psi)-c(u, 0, \psi)\Big)d\psi\\
&=&m _3 (-\infty).
 \eeQ        \qed
%%%%%%%%%%%%%%%%%%%%%%%%%%%%%%%%%%%%%%%%%%%%%%%%%%%%%%%%%%%%%%%%%%%%
\mysection{ADM and Bondi energy-momenta} \ls

In this section, we derive a formula relating the ADM total energy
and total linear momentum for a spacelike hypersurface at time $t
_0$ to the Bondi energy-momentum for a null hypersurface at
retarded time $u _0$ in non-radiative fields.
 \begin{thm}\label{ADM-Bondi-1}
Under {\bf Condition A}, {\bf Condition B} and {\bf Condition C},
the ADM total energy, the ADM total linear momentum of $N _{t _0}$
and the Bondi energy-momentum of null hypersurface $N _{u_0}$
satisfy
 \beQ
\mathbb{P} _\nu (t_0)&=&m _\nu (u _0)+\frac{1}{4\pi}\int
_{-\infty} ^{u _0} \int _{S ^2}\Big((c _{,0} )^2 +(d _{,0} )^2
\Big) n ^\nu dS du
 \eeQ
for $\nu$=$0$, $1$, $2$, $3$, and $\mathbb{E}(t_0)$ is denoted as
$\mathbb{P} _0 (t_0)$. In particular, if there is {\it news}, then
ADM total energy is always greater than the Bondi mass.
 \end{thm}
\pf By (\ref{u-deriv}), we obtain
 \beQ
M (u _0) &=&\lim _{u \rightarrow -\infty} M(u)+
\int _{-\infty} ^{u _0} M _{,0}du\\
&=&\lim _{u \rightarrow -\infty} M(u) +\int _{-\infty} ^{u _0}
\Big( (c _{,0}) ^2 +(d
_{,0}) ^2 \Big) du\\
& &+\frac{1}{2}\Big(l _{,2} +l \cot \theta +\bar l _{,3} \csc
\theta \Big) \Big| _{u=-\infty} ^ {u_0}.
 \eeQ
Then the theorem is a direct consequence of (\ref{u-Bondi-mass}),
(\ref{u-Bondi-momentum}), Theorem \ref{energy} and Theorem
\ref{momentum}. \qed
 \begin{rmk}\label{ADM-Bondi-2}
If we don't further assume {\bf Condition B}, then
 \beQ
\mathbb{E} (t_0)&=& m _0 (u _0)+\frac{1}{4\pi}\int _{-\infty} ^{u
_0} \int _{S ^2}\Big((c _{,0}) ^2 +(d _{,0}) ^2 \Big) dS du\\
 & &+\frac{1}{4\pi} \int _0 ^{2\pi}
\Big(c(u _0, 0, \psi)+c(u _0, \pi, \psi) \Big)d \psi.\\
 & &-\frac{1}{8\pi} \int _0 ^{2\pi}
\Big(c(-\infty, 0, \psi)+c(-\infty, \pi, \psi) \Big)d \psi,\\
 \mathbb{P} _1 (t_0)&=&m _1 (u _0)+\frac{1}{4\pi}\int _{-\infty}
^{u _0} \int _{S ^2}\Big((c _{,0}) ^2 +(d _{,0}) ^2 \Big) n ^1 dS
du,
\\
 \mathbb{P} _2 (t_0)&=&m _2 (u _0)+\frac{1}{4\pi}\int _{-\infty}
^{u _0} \int _{S ^2}\Big((c _{,0}) ^2 +(d _{,0}) ^2 \Big) n ^2 dS
du,\\
 \mathbb{P} _3 (t_0)&=&m _3 (u _0)+\frac{1}{4\pi}\int _{-\infty} ^{u
_0} \int _{S ^2}\Big((c _{,0}) ^2 +(d _{,0}) ^2 \Big) n ^3 dS du\\
& &+\frac{1}{4\pi}\int _0 ^{2 \pi} \Big(c (u _0, 0, \psi)-c(u _0,
\pi, \psi)\Big)d\psi.
 \eeQ
 \end{rmk}
 \begin{rmk}
If the spacetime $L ^{3,1}$ satisfies the dominant energy
condition, then the positive mass theorem \cite{SY1, SY2, SY3, W}
 \beQ
 \mathbb{E} \geq \Big( \mathbb{P} _1 ^2 +\mathbb{P}
_2 ^2 +\mathbb{P} _3 ^2\Big) ^{\frac{1}{2}}
 \eeQ
and Theorem \ref{ADM-Bondi-1} give rise to an inequality involving
the Bondi energy-momentum.
 \end{rmk}
%%%%%%%%%%%%%%%%%%%%%%%%%%%%%%%%%%%%%%%%%%%%%%%%%%%%%%%%%%%%%%%%%%%%
\mysection{Radiative fields}
\ls

It is a fundamental problem to find appropriate conditions of
asymptotically flatness at spatial infinity to include
gravitational radiation. Suggested by the Sommerfeld
electromagnetic boundary conditions, Trautman specified a class of
(noncovariant) boundary conditions for a spatially confined
gravitational source \cite{T}. The covariant formulation of the
Trautman boundary conditions was given by Papadopoulos and Witten
\cite{PW}.

The Bondi metric (\ref{bondi-metric}) gives rise to a class of
asymptotic flatness also. Instead of {\bf Condition C}, we assume
the metric has the following asymptotic behavior at spatial
infinity,
 \begin{description}
\item[Condition D] $\;\;\gamma \in \mathcal{C} _{\{1, 1, 1\}},\;\; \delta \in
\mathcal{C} _{\{1, 1, 1\}}, \;\;\beta \in \mathcal{C} _{\{2, 2,
2\}}, \;\;U \in \mathcal{C} _{\{2, 2, 2\}}, \;\;W \in \mathcal{C}
_{\{2, 2, 2\}}, \;\;V+r \in \mathcal{C} _{\{0, 0, 0\}}$.
 \end{description}
{\bf Condition D} implies, for $r$ sufficiently large,
 \beQ
& &\lim _{u \rightarrow -\infty}M=O\Big(1\Big), \lim _{u
\rightarrow -\infty}c=O\Big(1\Big),
\lim _{u \rightarrow -\infty}d=O\Big(1\Big),\\
& &\lim _{u \rightarrow -\infty}M _{,0}=O\Big(1\Big), \lim _{u
\rightarrow -\infty}c _{,0}=O\Big(1\Big),
\lim _{u \rightarrow -\infty}d _{,0}=O\Big(1\Big),\\
& &\lim _{u \rightarrow -\infty}M _{,A}=O\Big(1\Big), \lim _{u
\rightarrow -\infty}c _{,A}=O\Big(1\Big), \lim _{u \rightarrow
-\infty}d _{,A}=O\Big(1\Big)
 \eeQ
where $A,B=2, 3$. Physically, {\bf Condition D} might be an
interpretation of Sommerfeld's radiation condition at spatial
infinity: In \cite{BBM}, the authors found, in axi-symmetric
spacetime, this condition imply that $\gamma =\frac{f_1 (t-r)}{r}
+ \frac{f_2 (t-r)}{r} +\cdots$ as $r \rightarrow \infty$. As the
metric (\ref{bondi-metric}) behaves as a ``wave'', we may think
that $f_1, f_2, \cdots$ involve $\sin (t-r)$, $\cos(t-r)$, etc.
This is essential our motivation to introduce {\bf Condition D}.

Now we derive a formula between the ADM total energy and the Bondi
mass. This may be thought as the relation between them in
radiative fields. The relation between the ADM total linear
momentum $P _k$ and the Bondi momentum $m _k$ in radiative fields
requires much more dedicated computation. This question will be
addressed elsewhere.
 \begin{thm}\label{radi-energy}
Let $\mathbb{E} (t _0)$ be the ADM total energy of spacelike
hypersurface $N _{t_0}$ whose metric satisfies (\ref{metric}).
Under {\bf Condition A}, {\bf Condition B} and {\bf Condition D},
we have
 \beQ
\mathbb{E} (t _0) = m _0 (-\infty) +\frac{1}{4\pi}\lim
_{u\rightarrow -\infty} \int _0 ^{\pi} \int _0 ^{2\pi} \Big(c ^2
+d ^2 \Big) _{, 0} \sin\theta d\psi d\theta.
 \eeQ
 \end{thm}
\pf Note that (\ref{metric}) gives
  \beQ
g \big(\breve{e} _2, \breve{e} _2\big) +g \big(\breve{e} _3,
\breve{e} _3\big) &=&2 \cosh 2 \gamma \cosh 2 \delta \\
&=&2+4\Big(\gamma ^2 +\delta ^2 \Big) +O\Big(\frac{1}{r ^4}\Big)\\
&=&2+\frac{4\big(c ^2 +d ^2\big)}{r ^2}+O\Big(\frac{1}{r ^3}\Big).
  \eeQ
Hence, as $r \rightarrow \infty$, (or $u \rightarrow -\infty$),
 \beQ
\breve {\nabla } ^j g\big(\breve{e} _1, \breve{e} _j\big) -\breve
{\nabla } _1 tr _{g _0} \big(g\big)
 &=&\breve {e} _j \Big(g\big(\breve{e} _1, \breve{e} _j\big)\Big)
 -\breve{e} _1 tr _{g _0} \big(g\big)\\
 & &-g\big(\breve{e} _j, \breve{e} _i\big)\breve {\omega} _{i1}
 \big(\breve{e} _j\big) -g\big(\breve{e} _1, \breve{e} _i\big)\breve {\omega} _{ij}
 \big(\breve{e} _j \big)\\
 &=&\frac{4M}{r ^2} -\frac{l _{,2}}{r ^2} -\frac{l \cot \theta }{r
 ^2}-\frac{\bar l _{,3}}{r ^2 \sin \theta}\\
 & & -4\breve{e} _1\Big(\frac{c ^2 +d ^2}{r ^2} \Big)+O\Big(\frac{1}{r
 ^3}\Big)\\
 &=&\frac{4M}{r ^2} -\frac{l _{,2}}{r ^2} -\frac{l \cot \theta }{r
 ^2}-\frac{\bar l _{,3}}{r ^2 \sin \theta}\\
 & & +\frac{4\big(c ^2 +d ^2\big) _{,0}}{r ^2}+O\Big(\frac{1}{r
 ^3}\Big).
 \eeQ
This gives rise to Theorem \ref{radi-energy}.     \qed

%%%%%%%%%%%%%%%%%%%%%%%%%%%%%%%%%%%%%%%%%%%%%%%%%%%%%%%%%%%%%%%%%%%%
\mysection{Appendix}
\ls

The asymptotic expansion of the Christoffel symbols of
(\ref{metric}) under {\bf Condition A}, {\bf Condition B} and {\bf Condition C}:
 \beQ
\Gamma _{11} ^1 &=&-\frac{M +r M _{,0}}{r ^2} + O\Big(\frac{1}{r ^3}\Big),\\
\Gamma _{12} ^1 &=&\frac{l + M _{,2}}{r} + O\Big(\frac{1}{r ^2}\Big),\\
\Gamma _{13} ^1 &=&\frac{l \sin \theta +M _{,3}}{r} + O\Big(\frac{1}{r ^2}\Big),\\
\Gamma _{22} ^1 &=&-r + O\Big(1\Big),\\
\Gamma _{23} ^1 &=&\frac{1}{2}\Big(-\bar l _{, 2} \sin \theta
+\bar l \cos \theta
                   +l _{, 3} -2 d \sin \theta +2 r d _{,0} \sin \theta \Big)
                   + O\Big(\frac{1}{r}\Big),\\
\Gamma _{33} ^1 &=&-r \sin ^2 \theta + O\Big(1\Big),\\
\Gamma _{11} ^2 &=&-\frac{M _{, 2} -r l _{,0}}{r ^3} + O\Big(\frac{1}{r ^4}\Big),\\
\Gamma _{12} ^2 &=&\frac{1}{r} + O\Big(\frac{1}{r ^2}\Big),\\
\Gamma _{13} ^2 &=&-\frac{-l _{, 3} +\big(\bar l \sin \theta \big)
_{, 2}
                   -2\big(d + r d _{,0}\big)\sin \theta}{2r ^2}
                   + O\Big(\frac{1}{r ^3}\Big),\\
\Gamma _{22} ^2 &=&-\frac{l-c _{, 2}}{r} + O\Big(\frac{1}{r ^2}\Big),\\
\Gamma _{23} ^2 &=&-\frac{2d \cos \theta -c _{, 3}}{r}
                   + O\Big(\frac{1}{r ^2}\Big),\\
\Gamma _{33} ^2 &=&-\sin \theta \cos \theta+
O\Big(\frac{1}{r}\Big),\\
\Gamma _{11} ^3 &=&-\frac{M _{, 3} -r \bar l _{,0}}{r ^3 \sin ^2
\theta} +
O\Big(\frac{1}{r ^4}\Big),\\
\Gamma _{12} ^3 &=&-\frac{-l _{, 3} +\big(\bar l \sin \theta \big)
_{, 2}
                   +2r d _{,0}}{2r ^2 \sin ^2 \theta}+ O\Big(\frac{1}{r ^3}\Big),\\
\Gamma _{13} ^3 &=&\frac{1}{r}
                   + O\Big(\frac{1}{r ^2}\Big),\\
\Gamma _{22} ^3 &=&-\frac{\bar l \sin \theta -2d _{, 2} \sin
\theta -
                    2d \cos \theta +c _{, 3}}{r \sin ^2 \theta}
                   + O\Big(\frac{1}{r ^2}\Big),\\
\Gamma _{23} ^3 &=&\frac{\cos \theta }{\sin \theta} +O\Big(\frac{1}{r}\Big),\\
\Gamma _{33} ^3 &=&-\frac{\bar l \sin \theta +c _{, 3} -2d \cos
\theta }{r} +O\Big(\frac{1}{r ^2}\Big).
 \eeQ

{\footnotesize {\it Acknowledgements.} {The author is indebted to
Yun Kau Lau for pointing out to him the works of Ashtekar and
Magnon-Ashtekar, Rizzi and Hayward and many useful conversations,
and to Wenling Huang for her careful reading of the manuscript and
many valuable suggestions. This work was partially done when the
author visited the Center of Mathematical Sciences, Zhejiang
University and he would like to thank the center for its
hospitality. The research is partially supported by National Natural
Science Foundation of China under grant No. 10231050 and the
Innovation Project of Chinese Academy of Sciences. }


\begin{thebibliography}{99}
\bibitem {ADM} S. Arnowitt, S. Deser, C. Misner, {\it Coordinate
invariance and energy expressions in general relativity}, Phys.
Rev. 122(1961), 997-1006.
\bibitem {AH} A. Ashtekar, R. Hansen, {\it A unified treatment of
null and spatial infinity in general relativity. I. Universal
structure, asymptotic symmetries, and conserved quantities at
spatial infinity}, J. Math. Phys. 19(1978), 1542-1566.
\bibitem {AM} A. Ashtekar, A. Magnon-Ashtekar, {\it Energy-momentum
in general relativity}, Phys. Rev. Lett. 43(1979), 181-184.
\bibitem {B} R. Bartnik, {\it Bondi mass in the NQS gauge}, Class.
Quantum Grav. 21(2004) S59-S71.
\bibitem {BBM} H. Bondi, M. van der Burg, A. Metzner, {\it
Gravitational waves in general relativity VII. Waves from
axi-symmetric isolated systems}, Proc. Roy. Soc. London A
269(1962), 21-52.
\bibitem {CK} D. Christodoulou, S. Klainerman,
{\it The global nonlinear stablity of Minkowski space}, Princeton
Math. Series 41, Princeton Univ. Press, Princeton, 1993.
\bibitem {CJK} P. Chru\'sciel, J. Jezierski, J. Kijowski,
{\it Hamiltonian field theory in the radiating regime}. Lecture
Notes in Physics. Monographs, 70. Springer-Verlag, Berlin, 2002.
\bibitem {CJM} P. Chru\'sciel, J. Jezierski, M. MacCallum,
{\it Uniqueness of the Trautman-Bondi mass}, Phys. Rev. D58(1998),
084001.
\bibitem {CMS} P. Chru\'sciel, M. MacCallum, D. Singleton,
{\it Gravitational waves in general relativity. XIV. Bondi
expansions and the "polyhomogeneity" of $Scr\; I$.} Philos. Trans.
Roy. Soc. London A 350(1995), 113--141.
\bibitem {FK} H. Friedrich, J., Kannar, {\it Bondi-type systems near
space-time infinity and the calculation of the NP-constants}, J.
Math. Phys. 41(2000), 2195-2232.
\bibitem {H} S. Hayward, {\it Spatial and null infinity via advanced
and retarded conformal factors}, Phys. Rev. D68(2003), 104015.
\bibitem {HZ} W-l. Huang, X. Zhang, {\it On the relation between ADM
and Bondi energy-momenta III -- perturbed radiative spatial
infinity}, in preparation.
\bibitem {PR} R. Penrose, W. Rindler {\it Spinors and space-time},
Vols. I, II, Cambridge University Press 1984, 1988.
\bibitem {PT} T. Parker, C. Taubes, {\it On Witten's proof of the
positive energy theorem}, Commun. Math. Phys. 84(1982), 223-238.
\bibitem {PW} D. Papadopoulos, L. Witten, {\it Covariant boundary
conditions in gravitational radiation theory: A new covariant
definition of spatially asymptotically flat spacetimes}, Phys.
Rev. D23(1981), 267-271.
\bibitem {R} A. Rizzi, {\it Angular momentum in general relativity:
The definition at null infinity includes the spatial definition as
a special case}, Phys. Rev. D63(2001), 104002.
\bibitem {S} R. Sachs, {\it Gravitational waves in general
relativity VIII. Waves in asymptotically flat space-time}, Proc.
Roy. Soc. London, A 270(1962), 103-126.
\bibitem {SY1} R. Schoen, S.T. Yau, {\it On the proof of the
positive mass conjecture in general relativity}, Commun. Math.
Phys. 65(1979), 45-76.
\bibitem {SY2} R. Schoen, S.T. Yau, {\it The energy and the
linear momentum of spacetimes in general relativity}, Commun.
Math. Phys. 79(1981), 47-51.
\bibitem {SY3} R. Schoen, S.T. Yau, {\it Proof of the positive
mass theorem. II}, Commun. Math. Phys. 79(1981), 231-260.
\bibitem {T} A. Trautman, {\it Radiation and boundary conditions in
the theory of gravitation}, Bull. Acad. Pol. Sci., Ser. Sci. Math.
VI 407(1958)407-412.
\bibitem {V1} J. Valiente Kroon, {\it Early radiative properties of the
developments of time-symmetric conformally flat initial data},
Class. Quantum Grav. 20(2003)L53-L59.
\bibitem {V2} J. Valiente Kroon, {\it Does asymptotic simplicity allow
for radiation near spatial infinity}, Commun. Math. Phys.
251(2004), 211-234.
\bibitem {vdB} M. van der Burg, {\it Gravitational waves in general
relativity IX. Conserved quantities}, Proc. Roy. Soc. London A
294(1966), 112-122.
\bibitem {W} E. Witten, {\it A new proof of the positive energy
theorem}, Commun. Math. Phys. 80(1981), 381-402.
\bibitem {Z1} X. Zhang, {\it Angular momentum and positive mass
theorem}, Commun. Math. Phys. 206(1999), 137-155.
\bibitem {Z2} X. Zhang, {\it Remarks on the total angular momentum
in general relativity}, Commun. Theore. Phys., 39(2003), 521-524.
\bibitem {Z3} X. Zhang, {\it A definition of total energy-momenta
and the positive mass theorem on asymptotically hyperbolic
3-manifolds I}, Commun. Math. Phys., 249(2004), 529-548.
\end{thebibliography}
\end{document}